\documentclass[12pt]{article}
\usepackage{graphicx}

\begin{document}
\title {Colour entangled orphan quarks and dark energy from cosmic QCD phase 
transition} 
\date{}
\maketitle
\vspace{-1in}
\author{\begin{center}{Shibaji Banerjee$^1$, 
Abhijit Bhattacharyya$^2$,
Sanjay K. Ghosh$^3$,\\
Ernst-Michael Ilgenfritz$^4$, 
Sibaji Raha$^{3}$
Bikash Sinha$^5$,  
\footnote{Corresponding Author}
\footnote{Electronic Mail :
bikash@veccal.ernet.in}, 
Eiichi Takasugi$^6$ and Hiroshi Toki$^4$}
\end{center}
$^1$Physics Department, St. Xavier's College, 
30, Park Street, Kolkata 700016, INDIA \\
$^2$Physics Department, Scottish Church College, 1 \& 3, Urquhart Square, Kolkata 700006, INDIA \\
$^3$Physics Department, Bose Institute, 93/1, A. P. C. Road, 
Kolkata 700009, INDIA \\
$^4$Research Center for Nuclear Physics, Osaka University, Ibaraki, 
Osaka 567-0047, JAPAN \\
$^5$ Saha Institute of Nuclear Physics, 1/AF, Bidhannagar, Kolkata - 700 064, INDIA \& 
Speaker, QM'05 \\
$^6$ Graduate School of Physics, Osaka University, Toyonaka, Osaka 560-0043, JAPAN}
\begin{abstract}
The present day astrophysical observations indicate that the universe is composed of a large amount 
of dark energy (DE) responsible for an accelerated expansion of the universe, along with a 
sizeable amount of cold dark matter (CDM), responsible for structure formation. 
The explanations for the origin or the nature of both CDM and DE seem to require ideas beyond 
the standard model of elementary particle interactions. Here we show that 
CDM and DE both can arise from the standard principles of strong interaction physics and 
quantum entanglement. 
\end{abstract}

During the past few decades, several accurate 
astrophysical measurements have been carried out and the large amount of data collected makes it
possible to test the existing different cosmological models. 
Based on the knowledge gleaned so far, the present consensus \cite{1,2,3} is that the standard 
model of cosmology, comprising the Big Bang and a flat universe is correct. The Big Bang 
Nucleosynthesis \cite{4} (BBN), which forms one of the basic tenets of the standard model, 
shows that baryons can at most contribute \( \Omega_B \) ( \( \equiv \frac{\rho_B}{\rho_c} \), 
\( \rho_c \) being the present value of closure density \( \sim 10^{-47} \) Gev$^4$ ) 
\( \sim \)0.04, whereas structure formation studies require that the total (cold) matter 
density should be \( \Omega_{CDM} \sim 0.23 \). Matter contributing to CDM is characterized 
by a dust-like equation of state, pressure \( p \approx \) 0 and energy density \( \rho > 0 \) 
and is responsible for clustering on galactic or supergalactic scales. Dark energy (DE), 
on the other hand, is smooth, with no clustering features at any scale. It is required 
to have an equation of state \( p = w\rho \) where \( w< 0 \) (ideally \( w = -1 \)), 
so that for a positive amount of dark energy, the resulting negative pressure would 
facilitate an accelerated expansion of the universe, evidence for which has recently 
become available from the redshift studies of type IA supernovae \cite{3,5}. 
For a flat universe \( \Omega \sim 1 \), \( \Omega_{DE} \sim 0.73 \) \cite{6} 
implies that \( \rho_{DE} \) today is of the order of \( 10^{-48} \) GeV$^4$.

The BBN limit of 
\( \Omega_B \sim 0.04 \) has led to the argument that, CDM cannot be 
baryonic; as a result, various exotic possibilities, all beyond the standard model 
\( SU(3)_c \times SU(2) \times U(1) \) of particle interactions, have been suggested. 
The situation is even more complicated for DE. The most natural explanation for DE 
would be a vacuum energy density, which {\it a priori} would have the correct 
equation of state (\( w = -1 \)). The difficulty associated with such a possibility 
is the fact that for any known (or conjectured) type of particle interaction, 
the vacuum energy density scale turns out to be many orders of magnitude larger than the 
present critical density. 
In this paper, we show that it is possible to understand the nature of CDM and DE  
within the standard model \( SU(3)_c \times SU(2) \times U(1) \). 
They can both arise from the same process of 
the cosmic quark-hadron phase transition occurring during the microsecond epoch after 
the Big Bang, provided we admit the existence of quantum entanglement for a strongly 
interacting system. 

In this paper we assume a first order cosmic QCD phase transition. Other 
crucial {\it ans\"atze} in our scenario include 
that the universe is overall colour neutral at all times,
baryogenesis is complete substantially before the QCD 
transition epoch and that the baryon number is an integer. Together 
with the assumption of quantum entanglement, the above picture is capable of 
explaining both dark matter and dark energy. Here we will restrict ourselves
mainly to the case of dark energy. For details see \cite{15b}.

At temperatures higher than the critical temperature \( T_c \), 
the coloured quarks and gluons are in a thermally equilibrated state in the 
perturbative vacuum (the quark-gluon plasma). The total colour of the universe 
is neutral (i.e., the total colour wave function of the universe is a singlet). 
It is strongly believed from QCD analysis that this thermally equilibrated phase 
of quarks and gluons exhibits collective plasma-like behaviour, referred to in 
the literature as the Quark-Gluon Plasma or QGP. Then, as \( T_c \) is reached and 
the phase transition starts, bubbles of the hadronic phase, 
begin to appear in the quark-gluon plasma, grow in size and form an infinite chain 
of connected bubbles (the percolation process). At this stage, the ambient universe 
turns over to the hadronic phase. Within this 
hadronic phase, the remaining high temperature quark phase gets trapped in large 
bubbles. As is well known, this process 
is associated with a fluctuation in the temperature around \(T_c\); the bubbles of 
the hadronic phase can nucleate only when the temperature falls slightly below \(T_c\). 
The released latent heat raises the temperature again and so on. It is thus fair to 
assume that the temperature of the universe remains around \(T_c\) at least upto 
percolation.

The net baryon number contained in these 
Trapped False Vacuum Domains (TFVD) could be many orders of magnitude larger than 
that in the normal hadronic phase and they could constitute the absolute ground 
state of strongly interacting matter \cite{9}. The larger TFVDs with baryon number 
 \( \sim 10^{42-44} \), which are stable and separated by large distances 
(\( \sim \)300 m), would have a bearing on the
dark matter content of the universe \cite{12}. 
In all these considerations, it has been tacitly 
assumed that in a many-body system of quarks and gluons, colour is 
averaged over, leaving only a statistical degeneracy factor for thermodynamic 
quantities. Here we argue that such simplification may have led us 
to overlook a fundamentally important aspect of strong interaction physics in cosmology. 

Let us now elaborate on the situation in some detail. In the QGP, 
all colour charges are neutralised within the corresponding Debye length, 
which turns out to be \( \sim {1 \over{g_s(T) T}} \), where \(g_s\) is the 
strong coupling constant. For Debye length smaller than a typical 
hadronic radius, hadrons cannot exist as bound states of coloured 
objects. The lifetime of the QGP may be roughly estimated by the temperature 
when the Debye screening length becomes larger than the typical hadronic radius, 
and formation of hadrons as bound states of coloured objects becomes possible. 

The size of the universe being many orders of magnitude larger than the Debye 
length, the requirement of overall colour neutrality is trivially satisfied.  
Another condition required for the existence of the QGP would be the occurrence 
of sufficient number of colour charges within the volume characterised by the 
Debye length; otherwise the collective behaviour responsible for the 
screening would not be possible. For the cosmic QGP, these conditions are 
satisfied till temperatures \( \sim \) 100 - 200 MeV, the order of magnitude 
for the critical temperature for QCD. 

So one would argue that all the physics of the 
QGP is contained within a Debye length. A quick estimate would show that upto
\(T_c\), the Debye length is less than a fermi and the total number of colour 
charges (including quarks, antiquarks and gluons) within the corresponding 
volume is greater than 10. Note, however, our emphasis on the second ansatz 
above, that the baryogenesis has already taken place much before \(T_c\) 
is reached and the value of \({{n_b} \over {n_\gamma}} \sim 10^{-10} \) 
has already been established. Firstly, it has to be realised that this 
baryon number is carried in quarks at this epoch, so that the net quark 
number (\( N_q - N_{\bar{q}} \)) in the universe 
at all times is an exact multiple of 3. The net quark number within a 
Debye volume then turns out to be \(\sim 10^{-9}\). Thus, to ensure 
overall colour neutrality and an integer baryon number, one must admit 
of long-range correlations beyond the Debye length in the QGP, the 
quantum entanglement property \cite{15}, which was identified as an 
essential feature of quantum mechanics by Schr\"odinger in as early as 
1927 but was experimentally established only recently. 

Let us now consider the process of the cosmic quark-hadron phase transition 
from the quantum mechanical standpoint of colour confinement. As already 
mentioned, the colour wave function of the entire universe prior to the 
phase transition must be a singlet, which means it cannot be factorized 
into constituent product states; the wave functions of all coloured 
objects are completely entangled \cite{15} in a quantum mechanical 
sense. This also ensures the integer baryon number condition \cite{15b}.
In such a situation, the universe is 
characterized by a vacuum energy corresponding to the perturbative 
vacuum of Quantum Chromodynamics (QCD). As the phase transition proceeds, 
locally colour neutral configurations (hadrons) arise, resulting in gradual 
decoherence of the entangled colour wave function of the entire universe. 
This amounts to a proportionate 
reduction in the perturbative vacuum energy density, which goes into providing 
the latent heat of the transition, or in other words, the mass and the kinetic 
energy of the particles in the non-perturbative (hadronic) phase 
(the vacuum energy of the non-perturbative 
phase of QCD is taken to be zero). In the quantum mechanical 
sense of entangled wave functions, the end of the quark-hadron transition would 
correspond to complete decoherence of the colour wave function of the universe; 
the entire vacuum energy would disappear as the perturbative vacuum would be 
replaced by the non-perturbative vacuum. 

The earlier discussions imply that in order for the TFVDs to be stable physical objects, 
they must be colour neutral. This is synonymous with the requirement 
that they all have integer baryon numbers, i.e., at the moment of 
formation each TFVD has net quark numbers in exact multiples of 3. 
For a statistical process, this is, obviously, most unlikely and 
consequently, most of the TFVDs would have some residual colour 
at the percolation time. 
Then, on the way to becoming colour singlet 
they would each have to shed one or two coloured quarks.

Thus, at the end of the cosmic QCD phase transition 
there would be a few coloured quarks, separated by spacelike distances. 
Such a large separation, apparently against the 
dictates of QCD, is by no means unphysical. The separation of coloured TFVDs 
occurs at the temperature \( T_c \), when the effective string tension is zero, 
so that there does not exist any long range force. By the time the TFVDs have 
evolved into colour neutral configurations, releasing the few orphan coloured 
quarks in their immediate vicinity, the spatial separation between these 
quarks is already too large to allow strings to develop between them; 
see below. (Such a situation could not 
occur in the laboratory searches for quark-gluon plasma through 
energetic heavy ion collisions as the spatial extent of the system 
is \( \sim \)few fermi and the reaction takes place on strong 
interaction time scales.) Therefore, the orphan quarks must remain in isolation. 
In terms of the quantum entanglement and decoherence of the colour wave function, 
this would then mean that their colour wave functions must still remain entangled 
and {\it a corresponding amount of the perturbative vacuum energy would persist 
in the universe}. In this sense, the orphan quarks are definitely not in asymptotic 
states and no violation of colour confinement is involved. If we naively assume 
that the entanglement among these orphan quarks imply collective behaviour 
like a plasma, then one can estimate the corresponding Debye screening length 
for this very dilute system of quarks, which would still be governed primarily 
by the temperature. At temperatures of \( \sim \) 100 MeV, the scale being 
much smaller than the mutual separation between the orphan quarks, the formation 
of bound states of orphan quarks would be impossible. 

Presently, it is not possible to calculate this persisting, or even the full, 
perturbative vacuum energy, from first principles in QCD. For the latter 
quantity, one may adopt the phenomenological Bag model \cite{16} of 
confinement, where the Bag parameter B (\(\sim (145 MeV)^4 \)) is the 
measure of the difference between the perturbative and the non-perturbative 
vacua. Thus we can assume that at the beginning of the phase transition, 
the universe starts out with a vacuum energy density B, which gradually 
decreases with increasing decoherence of the entangled colour wave function. 
A natural thermodynamic measure of the amount of entanglement during the 
phase transition could be the volume fraction 
(\( f_q \equiv V_{colour}/V_{total} \)) of the coloured degrees of freedom; 
at the beginning, \(f_q\) is unity, indicating complete entanglement, 
while at the end, very small but finite entanglement corresponds to a 
tiny but non-zero \(f_q\) due to the coloured quarks. Accordingly, the 
amount of perturbative vacuum energy density in the universe at any 
time is the energy density B times the instantaneous value of \(f_q\); 
within the scenario discussed above, the remnant perturbative vacuum 
energy at the end of the QCD transition would just be B \( \times f_{q,O}\), 
where \(f_{q,O}\) is due solely to the orphan quarks. An order of magnitude 
estimate for \(f_{q,O}\) can be carried out in the following straightforward 
manner. On the average, each TFVD is associated with 1 orphan quark so that 
the number \( N_{q,O} \) of orphan quarks within the horizon volume at any 
time is about the same as the number \( N_{TFVD} \) of TFVDs therein. It is 
well known from the study of percolating systems \cite{18} that percolation 
is characterized by a critical volume fraction \(f_c \sim \) 0.3 of the high 
temperature phase. So in the present case, \(f_q \) in the 
form of TFVD-s would be \( \sim \) 0.3. Following the ansatz of Witten \cite{9} 
that the most likely length scale for a TFVD is a few cm, one can estimate 
\(N_{TFVD}\) (and hence \(N_{q,O}\)) within the horizon at the percolation 
time of about 100 \(\mu\)sec \cite{12} to be about 10\(^{18-20}\). The inter-TFVD 
separation comes out to be \( \sim \) 0.01 cm at that time. (It is obvious that 
the orphan quarks, separated by distances of 0.01 cm, cannot develop colour strings 
between them, even if there is some non-zero string tension generated at 
temperatures slightly lower than \(T_c\).) Then, if we naively associate an 
effective radius of \(\sim 10^{-14} cm\) (estimated from 
\( \sigma_{qq} = \frac{1}{9} \sigma_{pp}~;~~ \sigma_{pp} \sim 20 mb \) ) 
with each orphan quark, we obtain \( f_{q,O} \sim N_{q,O} \times (v_{q,O}/V_{total}) 
\sim 10^{-42} - 10^{-44}\) (where $v_{q,O}$ is the effective volume of an 
orphan quark), so that the residual pQCD vacuum energy comes out to be in 
the range 10\(^{-46}\) to 10\(^{-48} GeV^4\) \cite{15b}, just the amount of DE. 

To conclude, we have shown that the existence of DE (along with CDM) can be explained 
entirely within 
the standard model of particle interactions, without invoking any exotic assumption.
It is remarkable that this simple picture gives quantitatively 
correct amount without any fine-tuning of parameters.
Moreover, the picture presented here satisfies, in our 
opinion, the most stringent criterion of a scientific theory, 
that of naturalness.

SR would like to thank the Research Center for Nuclear Physics, Osaka University 
for their warm hospitality during his sojourn there.


\begin{thebibliography}{999} 

\bibitem{1} T. Tegmark, {\it Science} {\bf{296}}, 1427-1433 (2002). 

\bibitem{2} S. Sarkar, Proc. EPS Int.Conf. on High Energy Physics, Budapest, 2001 (Horvath, D., 
Levai, P. and Patkos, A., eds.) JHEP Proceedings Section, PrHEP-hep2001/299.

\bibitem{3} B. Leibundgut, {\it Ann. Rev. Astron. Astrophys.} {\bf{39}}, 67-98 (2001).

\bibitem{4} C. J. Copi, D. N. Schramn, and M. S. Turner, {\it Science} {\bf{267}}, 192-199 (1995).

\bibitem{5} R. P. Kirshner, {\it Science} {\bf{300}}, 1914-1918 (2003).

\bibitem{6} H. V. Peiris et al., {\it astro-ph}/0302225. 

\bibitem{15b} S. Banerjee et al., {\it Phys. Lett.} {\bf {B 611}}, 27-33 (2005).

\bibitem{9} E. Witten, {\it Phys. Rev.} {\bf{D30}}, 272-285 (1984).

\bibitem{12} S. Banerjee  et al., {\it Mon. Not. R. Astron. Soc.} {\bf{340}}, 284-288 (2003).

\bibitem{15} W. K. Wootters, {\it Phil. Trans. R. Soc. Lond.} {\bf{A356}}, 1717-1731 (1998). 

\bibitem{16} A. Chodos,  R. L Jaffe, K. Johnson, C. B. Thorn and V. F. Weisskopf, {\it Phys. Rev.} 
{\bf{D9}}, 3471-3495 (1974).

\bibitem{18} D. Stauffer, {\it Phys. Rep. } {\bf 54}, 1-74 (1979).

\end{thebibliography}
\end{document}